\documentstyle[prb,aps]{revtex}
\begin{document}
\title{Ferromagnetism and Canted Spin Phase in AlAs/GaMnAs Single Quantum Wells:
Monte Carlo Simulation}
\author{M. A. Boselli}
\address{Instituto de F\'\i sica, Universidade do Estado do Rio de Janeiro\\
Rua S\~{a}o Francisco Xavier 524, 22.500-013 Rio de Janeiro, R.J., Brazil}
\author{A. Ghazali}
\address{Groupe de Physique des Solides, Universit\'{e}s Paris 7 et Paris 6\\
Tour 23, 2 Place Jussieu, F-75 125 Paris Cedex 05, France}
\author{I. C. da Cunha Lima}
\address{Instituto de F\'\i sica, Universidade do Estado do Rio de Janeiro\\
Rua S\~{a}o Francisco Xavier 524, 22.500-013 Rio de Janeiro, R.J., Brazil and}
\maketitle
\date{\today}

\begin{abstract}
The magnetic order resulting from a confinement-adapted
Ruderman-Kittel-Kasuya-Yosida indirect exchange between magnetic moments in
the metallic phase of a AlAs$/$Ga$_{1-x}$Mn$_{x}$As quantum well is studied
by Monte Carlo simulation. This coupling mechanism involves magnetic moments
and carriers (holes), both coming from the same Mn$^{2+}$ ions. It leads to
a paramagnetic, a ferromagnetic, or a canted spin phase, depending on the
carrier concentration, and on the magnetic layer width. It is shown that
high transition temperatures may be obtained.
\end{abstract}
\pacs{75.70.-b, 75.10.-i, 75.70.Cn}

\section{Introduction}

\label{intro}

During the last decade, due to the advances in the control of materials
growth, and also in the techniques of characterization, a new interest arose
in the study of the magnetic order in layered materials. This area is not
restricted to magnetism in metals, but it also includes the study of
magnetic semiconductor pseudo-binary alloys like A$_{1-x}$M$_{x}$B, where M
stands for a magnetic ion. These alloys are called Diluted Magnetic
Semiconductors (DMS). \cite{dms1,dms2}

Recently some groups \cite{van1,van2,oiwa1,oiwa2,matsukura,ohno1,ohno2}
succeeded in producing homogeneous samples of Ga$_{1-x}$Mn$_{x}$As alloys
with $x$ up to $7\%$ using low temperature ($200-300^{o}$ C) MBE
techniques. Mn is a transition metal having its $3d$ level partially filled
with five electrons, in such a way that it carries a magnetic moment of $%
5\hbar /2$, according to the Hund's rule. In the insulating phase, as in
II-VI DMS, two Mn$^{2+}$ ions occupying the nearest neighbor positions are
assumed to interact with each other {\it via} a super-exchange mechanism,
resulting in an anti-ferromagnetic ordering of their magnetic moments. In
the fcc alloys, these interactions are frustrated, establishing the
possibility of settling a spin-glass phase at low temperature. A double
exchange mechanism which might stabilize a ferromagnetic coupling between
the Mn ions in III-V DMS has been suggested by Akai, \cite{akai} but has not
been confirmed in the EPR experiments performed by Szczytko et al, \cite
{scztw} who did not observe trace of neutral Mn, concluding that the double
exchange mechanism is not effective.

The possibility of having a DMS based on GaAs opens a wide range of
potential applications such as integrated magnetooptoelectronic devices.
Besides its practical importance, this kind of DMS introduces an interesting
problem: an Mn impurity in GaAs is an acceptor (it binds one hole), and
at the same time it carries a localized magnetic moment.
In the Ga$_{1-x}$Mn$_{x}$As alloy Mn is, in fact, a
strong $p$ dopant, \cite{van1,matsukura} the free
hole concentration reaching even $10^{20-21}cm^{-3}$. At small Mn
concentrations, the alloy is a paramagnetic insulator. As $x$ increases it
becomes ferromagnetic, going through a non-metal-to-metal transition for
higher concentrations, and keeping its ferromagnetic phase. For $x$ above $%
7\%$, the alloy becomes a ferromagnetic insulator. \cite{matsukura} In the
metallic phase, depending on the value of $x$, the temperature of the
ferromagnetic transition is observed in the range of 30-100 K, the highest
values observed in DMS. The ferromagnetic order in the metallic phase is
understood, at present, as resulting from the indirect exchange between the Mn$%
^{2+}$ ions due to the spin polarization of the hole gas.

The aim of this work is to study the magnetic order resulting of the
indirect exchange between magnetic moments in a GaAlAs$/$Ga$_{1-x}$ Mn$_{x}$%
As quantum well. A confinement-adapted Ruderman-Kittel-Kasuya-Yosida (RKKY)
\cite{rkky1,rkky2,rkky3,rkky4} mechanism is believed to be the most
important interaction in such systems, if sufficiently strong doping is
provided, as it is the case in metallic samples. It leads to an indirect
exchange coupling between Mn$^{2+}$ ions, mediated by carriers (holes),
which come from the same Mn$^{2+}$ ions.

This article is organized as follows. In Sec. \ref{rkky} we present the
calculation of the RKKY exchange for a confined Fermi gas in a semiconductor
heterostructure. For the sake of relating our results with other previous
ones, we explicitly separate our calculations as intra-subband and
inter-subband contributions. We emphasize that, in the quantum limit, i.e.,
when only the first subband is occupied, the intra-subband exchange is
factorized into a purely 2-D RKKY exchange times a form factor determined by
the architecture of the confining structure.\cite{luiz}

In Sec. \ref{moncar} a Monte Carlo simulation is performed to determine the
resulting magnetic phases and the relevant properties. Our calculations
reveal that a ferromagnetic order may occur in a single DMS quantum well
only beyond a minimum width of the magnetic layer, otherwise the sample is
paramagnetic. This is in keeping with recent experiments,\cite{haya} and is
a consequence of the need of a certain number of magnetic neighbors before a
ferromagnetic phase settles in. Depending on the well width and on the
effective two-dimensional carrier concentration, a canted phase can occur,
with a sizeable net low-temperature magnetization, $<S>/S_{max}$, and a well
behaved Edwards-Anderson order parameter $q$. The origin of the canted spin
phase is investigated by analyzing the parallel and the perpendicular
magnetizations, and the spin-spin correlation function components. The
magnetic susceptibilities are calculated in the
existing phases.

Finally, in Sec. \ref{comments} we summarize the results obtained, and comment
about the expected magnetic order in the structures analyzed.

\section{RKKY interaction in a quantum well}

\label{rkky}

The RKKY interaction between localized magnetic moments imbedded in a Fermi
gas is a well understood problem, since its early developments almost fifty
years ago. However, the new areas for experimental research brought into
evidence some theoretical problems concerning the RKKY interaction in low
dimensional systems, so far unexplored, such as purely 2-D and 1-D
arrangements of magnetic moments, interaction between magnetic layers,
magnetic moments in inhomogeneous electron gas, etc. \cite
{yafet,bergmann,bruno1,edw1,frank,bruno2,balten1,balten2,peeters} The
indirect exchange between localized magnetic moments in a quantum well
mediated by a Fermi gas has been addressed several times. Basically, it
deals with a confined electron (or hole) gas, therefore a
quasi-two-dimensional system, being locally polarized by magnetic moments
distributed in a layer. To our knowledge, Korenblit and Shender,\cite{koren}
were the first to obtain a closed expression to the equivalent of the RKKY
interaction in the limiting situation of a purely 2-D electron gas, although
Kittel\cite{rkky4} obtained a numerical solution to this question earlier.
Larsen \cite{larsen} derived an expression for a general dimensionality,
reproducing Korenblit and Shender results for d=2.
A detailed calculation to obtain a closed expression in
2-D was shown by B\'{e}al-Monod.\cite{beal} Gummich and da Cunha Lima
\cite{utiv} studied the indirect exchange between magnetic impurities in a
doped GaAs/AlAs quantum well in the diluted regime, obtaining a ferromagnetic
interaction. Finally, another expression for
a generic dimensionality has been derived by Aristov. \cite{aristov} The
approximation of the Fermi gas in a quantum well by a purely 2-D system is
seldom a reasonable choice. Helman and Baltensperger \cite{balten1,balten2}
treated the question of the polarization of an inhomogeneous electron gas in
several circumstances, emphasizing the roles of the confined and extended
states. The specific case of a DMS quantum well was addressed recently by
Dietl et al, \cite{dietl97} but they assumed that the magnetic moments
are spread all over the region allowed to the carriers, and in that case the
inter-subband contributions to the Curie-Weiss temperature cancel out in
a mean field approximation.

The interaction potential between a Fermi gas and a set of localized
magnetic moments at positions $\vec{R}_{i}$ is well described by the
Hund-type exchange potential:

\begin{equation}
H_{\text{ex}}=-I\sum_i\vec S_i\cdot \vec s(\vec r)\delta(\vec r- \vec R_i),
\label{hund}
\end{equation}
where $\vec S_i$ is the spin of the magnetic moment at position $\vec R_i$,
which will be treated as a classical variable, and $\vec s(\vec r) $ is the
spin operator of the fermion at $\vec r$. $I$ is the $sp-d$ interaction.\cite
{szcz} If $\hat \psi _\sigma (\vec r)$ and $\hat \psi _\sigma ^{\dagger }(%
\vec r)$ describe the fermion field operator for spin $\sigma $, then

\begin{equation}
s^z(\vec r)=\frac 12(\hat \psi _{\uparrow }^{\dagger }(\vec r)\hat \psi
_{\uparrow }(\vec r)-\hat \psi _{\downarrow }^{\dagger }(\vec r)\hat \psi
_{\downarrow }(\vec r)),  \label{sigz}
\end{equation}

\begin{equation}
s^{+}(\vec r)=\hat \psi _{\uparrow }^{\dagger }(\vec r)\hat \psi
_{\downarrow }(\vec r),  \label{sig+}
\end{equation}

\begin{equation}
s^{-}(\vec{r})=\hat{\psi}_{\downarrow }^{\dagger }(\vec{r})\hat{\psi}%
_{\uparrow }(\vec{r}),  \label{sig-}
\end{equation}
with the usual definitions of $s^{+}=s_{x}+is_{y}$, and $s^{-}=s_{x}-is_{y}$%
. Instead of free fermions in a 3-D space, the electrons and holes in a
ssemiconductor heterostructure are confined in the growth direction, assumed
to be the z-axis, due to the mismatch of the conduction and valence band
edges. Since they are free particles in the plane perpendicular to that
growth direction, i.e., in the plane parallel to the semiconductor
interfaces, their field operator are given by:

\begin{equation}
\hat{\psi}_{\sigma }(\vec{r})=\frac{1}{\sqrt{A}}\sum_{n,\vec{k}}e^{i\vec{k}.%
\vec{R}}\phi _{n}(z)\eta c_{n,\vec{k},\sigma },  \label{field}
\end{equation}
where $A$ is the normalization area, $\vec{k}$ is a wavevector in the plane (%
$x,y$), $\eta $ is the spin tensor, $\phi _{n}(z)$ is the envelope function
which describes the motion of the fermion in the $z$-direction, and $c_{n,%
\vec{k},\sigma }$ is the fermion annihilation operator for the state ($n,%
\vec{k},\sigma $). $\vec{R}$ represents a vector in the 2-D coordinates
plane ($x,y$). The usual RKKY perturbation calculation up to second order
leads to the correction on the ground state energy of the system formed by
the set of (classical) localized moments and the Fermi gas :

\begin{equation}
\delta E^{(2)}=\delta E_a^{\left( 2\right) }+\delta E_b^{\left( 2\right) },
\end{equation}
where

\begin{equation}
\delta E_{a}^{\left( 2\right) }=-\left( \frac{I}{2N}\right)
^{2}\sum_{i}\sum_{n,n^{\prime }}\mid \phi _{n}(z_{i})\mid ^{2}\mid \phi
_{n^{\prime }}(z_{i})\mid ^{2}S_{i}(S_{i}+1) \sum_{\vec{q}}\chi
^{n,n^{\prime}}(\vec{q}),  \label{self}
\end{equation}
and 
\begin{equation}
\delta E_{b}^{\left( 2\right) }=-\left( \frac{I}{2N}\right)
^{2}\sum_{j}\sum_{i\neq j}\sum_{n,n^{\prime }}\sum_{\vec{q}}2\mbox{Re}\left[
\phi _{n}^{\ast }(z_{i})\phi _{n^{\prime }}(z_{i})\phi _{n^{\prime }}^{\ast
}(z_{j})\phi _{n}(z_{j})e^{-i\vec{q}.(\vec{R}_{i}-\vec{R}_{j})}\right] \chi
^{n,n^{\prime }}(\vec{q})\vec{S}_{i}\cdot \vec{S}_{j}.  \label{exc}
\end{equation}
Eqs. (\ref{self}) and (\ref{exc}) are, respectively, the self-energy term
and the RKKY exchange in the form they assume for confined fermions. The
coordinates ($\vec{R}_{i},z_{i}$) describe the position of the impurity $i$
in the plane ($x,y$), and in the growth direction; $\vec q$ is a
two-dimensional wavevector. $\chi ^{n,n^{\prime }}(\vec{q})$ is the
equivalent to the Lindhard function:\cite{abrik,keld}

\begin{equation}
\chi ^{n,n^{\prime }}(\vec q)=\sum_{\vec k}\frac{\theta (E_F-\epsilon _{n,%
\vec k})-\theta (E_F-\epsilon _{n^{\prime },\vec k+\vec q})}{\epsilon
_{n^{\prime },\vec k+\vec q}-\epsilon _{n,\vec k}}.  \label{modlin}
\end{equation}
Eqs. (\ref{self}) and (\ref{exc}) are used to define the exchange
Hamiltonian:

\begin{equation}
H_{ex}=-\sum_{i,j}J_{ij}\vec S_i\cdot \vec S_j.  \label{excham}
\end{equation}
For $i\neq j$,

\begin{equation}
J_{ij}=\left( \frac{I}{2A}\right) ^{2}\sum_{n,n^{\prime }}\sum_{\vec{q}}2%
\mbox{ Re}\left[ \phi _{n}^{\ast }(z_{i})\phi _{n^{\prime }}(z_{i})\phi
_{n^{\prime }}^{\ast }(z_{j})\phi _{n}(z_{j})e^{-i\vec{q}.(\vec{R}_{i}-\vec{R%
}_{j})}\right] \chi ^{n,n^{\prime }}(\vec{q}).  \label{generj}
\end{equation}

\subsection{Intra-subband terms}

To our knowledge, complete calculations of Eq. (\ref{exc}) have only been
performed for intra-subband transitions, using different approaches. For the
sake of completeness, we will show how this is achieved in our treatment.
The contribution of a subband $n$ to the exchange reads:

\begin{equation}
J_{ij}^{(n)}=\left( \frac{I}{2A}\right) ^{2}\mid \phi _{n}(z_{i})\mid
^{2}\mid \phi _{n}(z_{j})\mid ^{2}\sum_{\vec{q}}2\cos (\vec{q}\cdot \vec{R}%
_{ij})\chi ^{n,n}(\vec{q}).  \label{intraexc}
\end{equation}

We observe that, in the so-called quantum limit, when only the first subband
($n=0$) is occupied, the difference between Eq. (\ref{intraexc}) and the
indirect exchange mediated by a 2D electron gas comes from the non-uniform
charge density in the confining direction $z$. Actually, in that case, Eq. (%
\ref{intraexc}) factorizes into a form factor ${\cal F}^{ij}$ and a purely
2-D exchange: \cite{luiz}

\begin{equation}
J_{ij}^{(0)}={\cal F}^{ij}J_{ij}^{(2-D)}.  \label{jluiz}
\end{equation}
where 
\begin{equation}
{\cal F}^{ij}=\mid \phi _{0}(z_{i})\mid ^{2}\mid \phi _{0}(z_{j})\mid ^{2}
\label{calf}
\end{equation}
and 
\begin{equation}
J_{ij}^{2-D}=\left( \frac{I}{2A}\right) ^{2}\sum_{\vec{q}}2\cos (\vec{q}%
\cdot \vec{R}_{ij})\chi ^{n,n}(\vec{q}).  \label{jij2d}
\end{equation}

It is easy to show, by using the dimensionless variables $x=kR_{ij}$ , and $%
y=qR_{ij}$, that the Fourier transform of the modified Lindhard function,
appearing in the summation in $\vec{q}$ at the rhs of Eq. (\ref{intraexc}),
becomes:

\begin{equation}
\chi ^{n}(R_{ij})=\frac{4m_{t}^{\ast }A^{2}}{\pi ^{3}\hbar ^{2}R_{ij}^{2}}%
\int_{0}^{\infty }dyyJ_{0}(y)\int_{0}^{k_{F}^{(n)}.R_{ij}}dxx\int_{0}^{\pi
/2}d\phi \frac{1}{y^{2}-4x^{2}\cos ^{2}\phi },
\end{equation}
where $\chi ^{n}(R_{ij})=\sum_{\vec{q}}2\cos (\vec{q}\cdot \vec{R}_{ij})\chi
^{n,n}(\vec{q})$. The transversal effective mass, $m_{t}^{\ast }$, is
assumed as isotropic in the plane parallel to the interfaces. As usual, $%
k_{F}^{(n)}=$ $\sqrt{2m_{t}^{\ast }(E_{F}-\epsilon _{n})}/\hbar $.
Performing the $\phi $ integral and changing variables again ($%
y/2x\rightarrow y$),

\begin{equation}
\chi ^{n}(R_{ij})=\frac{2m_{t}^{\ast }A^{2}}{\pi ^{2}\hbar ^{2}R_{ij}^{2}}%
\int_{0}^{k_{F}^{(n)}.R_{ij}}dxx\int_{1}^{\infty }dyJ_{0}(2xy)\frac{1}{\sqrt{%
y^{2}-1}}.  \label{phase1}
\end{equation}
The integral on $y$ is straightforward:\cite{grads}

\begin{equation}
\chi ^{n}(R_{ij})=-\frac{m_{t}^{\ast }A^{2}}{\pi \hbar ^{2}R_{ij}^{2}}%
\int_{0}^{k_{F}^{(n)}.R_{ij}}dxxJ_{0}(x)N_{0}(x).  \label{phase2}
\end{equation}
After performing the integral on $x$, Eq. (\ref{phase2}) results in:

\begin{equation}
\chi ^{n}(R_{ij})=-\frac{m_{t}^{\ast }A^{2}}{\pi \hbar ^{2}}%
k_{F}^{(n)2}[J_{0}(k_{F}^{(n)}R_{ij})N_{0}(k_{F}^{(n)}R_{ij})+J_{1}(k_{F}^{(n)}R_{ij})N_{1}(k_{F}^{(n)}R_{ij})].
\label{chiintra}
\end{equation}

This expression for the real space Lindhard function has been derived in a
different context, by several authors.\cite{koren,utiv,beal,aristov} The
final expression for the intra-subband exchange becomes:

\begin{eqnarray}
J_{ij}^{(n)} &=&-\left( \frac{I}{2}\right) ^{2}\frac{m_{t}^{\ast }}{\pi
\hbar ^{2}}k_{F}^{(n)2}\mid \phi _{n}(z_{i})\mid ^{2}\mid \phi
_{n}(z_{j})\mid ^{2}\times  \nonumber \\
&&[J_{0}(k_{F}^{(n)}R_{ij})N_{0}(k_{F}^{(n)}R_{ij})+J_{1}(k_{F}^{(n)}R_{ij})N_{1}(k_{F}^{(n)}R_{ij})].
\label{jintra}
\end{eqnarray}

\subsection{Inter-subband terms}

The contribution of the inter-subband terms cannot be expressed easily in a
closed form. Starting over from Eq. (\ref{generj}), and using the same
approach as in Ref. [13], we arrive to:

\begin{equation}
J_{ij}^{(n,n^{\prime })}=\left( \frac{I}{2}\right) ^{2}\frac{1}{\pi }%
\mbox{Re}\left[ \phi _{n^{\prime }}^{\ast }(z_{i})\phi _{n}(z_{i})\phi
_{n}^{\ast }(z_{j})\phi _{n^{\prime }}(z_{j})\right] \int_{0}^{\infty
}dqqF_{n,n^{\prime }}(q)J_{0}(qR_{ij}),  \label{stj}
\end{equation}
where we used

\begin{equation}
F_{n,n^{\prime }}(q)=\frac{m_{t}^{\ast }}{\hbar ^{2}}\frac{1}{\pi ^{2}}\int
d^{2}k\frac{q^{2}+\Delta _{n^{\prime },n}}{(q^{2}+\Delta _{n^{\prime
},n})^{2}-(2\vec{k}\cdot \vec{q})^{2}}\theta (\epsilon _{n^{\prime }}-E_{F}),
\label{deff}
\end{equation}
and $\Delta _{n^{\prime },n}=2m_{t}^{\ast }\cdot (E_{n^{\prime
}}-E_{n})/\hbar ^{2}$. The integral in Eq. (\ref{deff}) is, then,
straightforward:

\begin{equation}
F_{n,n^{\prime }}(q)=\frac{m_{t}^{\ast }}{2\pi \hbar ^{2}}(1-\frac{\Delta
_{n^{\prime },n}}{q^{2}})[1-\sqrt{1-(\frac{2k_{F}^{(n)}q}{q^{2}+\Delta
_{n^{\prime },n}})^{2}}\theta (q^{2}+\Delta _{n^{\prime
},n}-2qk_{F}^{(n)})]\theta (\epsilon _{n^{\prime}}-E_{F}).
\end{equation}

\section{Monte Carlo simulation: Magnetic ordering}

\label{moncar}

In order to determine the possible magnetic order in GaAs:Mn quantum wells,
we have performed extensive Monte Carlo simulations. Classical spins ${\bf S}%
_{i}$, randomly distributed on the cation sites with concentration $x$, are
assumed to interact through the RKKY exchange Hamiltonian defined by Eq. (%
\ref{excham}).

In the present work we have focused our attention on metallic single quantum
wells with the Mn concentration $x=5\%$, and we have neglected possible
(anti-)ferromagnetic interaction between the nearest neighbors and the next
nearest neighbors pairs. The RKKY exchange interaction derived in Sec. \ref
{rkky} is assumed to be effective within a cutoff radius which we have taken
as $R_{c}=4a$, $R_{c}=2a$, and $R_{c}=a$, where $a$ is the fcc lattice
parameter of GaAs. This makes the smallest value assumed for $R_{c}$ nearly
equal to the hole mean free path estimated from bulk transport measurements.
\cite{matsukura} The highest value, $R_{c}=2.2$nm, amounts to 3-4 values of
that mean free path. The consequences of the cutoff radius on the results
will be discussed below.

The calculation is performed in a finite box, whose axes are parallel to the
[100] directions. Its dimensions $\ $are $L_{x}=L_{y}$, and $L_{z}=Na/2$,
and $N$ is the number of DMS monolayers (ML) in the barrier. Periodic
boundary conditions are imposed in the $(x,y)$ plane. The lateral dimensions
are adjusted in such a way that the total number $N_{s}$ of spins is about
4400, for all samples with different $L_{z}$. The initial spin orientations
are randomly assigned. At a given temperature, the energy of the system due
to RKKY interaction is calculated, and the equilibrium state for a given
temperature is sought by changing the individual spin orientation according
to the Metropolis algorithm. \cite{diep} A slow cooling stepwise process is
accomplished making sure that the thermal equilibrium is reached at every
temperature. The resulting spin configuration is taken as the starting
configuration for the next step at a lower temperature.

For every temperature, the average magnetization $<M>$ and the
Edwards-Anderson (EA) order parameter $q$ are calculated. \cite{diep} The
latter is defined as 
\begin{equation}
q=\frac{1}{N}\sum_{i=1}^{N} \left( \sum_\alpha \left| \frac{1}{t}%
\sum_{t^{\prime }=t_{0}}^{t_{0}+t}{S}_{i \alpha}(t^{\prime }) \right|^2
\right)^{1/2},  \label{eadef}
\end{equation}
where $\alpha=x$, $y$ and $z$. In order to avoid spurious results in
obtaining the average over a large time interval $t$, a summation on $%
t^{\prime }$ is performed starting from a time $t_{0}$, when the system
already reached the thermal equilibrium.

In our calculations we used the value $N_{0}\beta =-1.2$ eV, taken from Ref. %
\onlinecite{okabayashi}, for which we obtain transition temperatures in good
agreement with the experimental data. Recently, the value $N_{0}\beta =-0.9$
eV has been obtained theoretically, \cite{bhatt} confirming the result of
Ref.\onlinecite{okabayashi}. The earlier estimate \cite{matsukura} $%
|N_{0}\beta |=3.3$ eV is probably too high.

Monte Carlo calculations have been performed in sixteen samples, as shown in
Table \ref{tab01}, the well widths varying from 25 {\AA } to 100 {\AA } (9
ML to 35 ML). In sample \#01, for a well width of 50 \AA\, and assuming $%
R_{c}=8$ ML, we tested the effect of a small hole concentration, making $p$
just 1\% of $x$. This amounts to $p \approx 1.1 \times 10^{19} cm^{-3}$. We
found that the spins can be arranged in a ferromagnetic phase even at this
carrier concentration. For that sample the calculation gives a transition
temperature near 27~K.

In Fig. \ref{fig01} the normalized average magnetization is shown for
samples \#02 to \#07 with different well widths, but with a fixed carrier
concentration of 10\% of $x$, i.e., $p \approx 1.1 \times 10^{20} cm^{-3}$,
and the same cutoff $R_{c}=8$ ML. For a very thin well (25 \AA ), we found
that the sample is paramagnetic. The EA order parameter for these samples is
shown in Fig. \ref{fig02}. It can be observed that, for sample \#02, there
is no phase transition. All the other curves in Fig. \ref{fig02} (samples
\#03 to \#07) are characteristic of an ordered phase. Raising the well width
(starting from 35 \AA ), the number of interacting neighbors increases, and
the sample shows successively a ferromagnetic phase (samples \#03, \#04, and
\#06) and a canted spin arrangement (samples \#5 and \#7). The value chosen
for the cutoff radius is larger than the first zero of the $J_{ij}^{\text{%
RKKY}}$, so antiferromagnetic interactions are turned on. With these choices
of $p$ and $R_{c}$, depending on the well width, the antiferromagnetic
interactions can settle a fraction of the Mn magnetic moments antiparallel.
This is the origin of the canted spins phase. The final average
magnetizations in Fig. \ref{fig01}  at $T=0$ K are only a fraction of
the maximum magnetization, around 60\% for sample \#5, and 70\% for sample \#7.
The transition temperatures were found in the range of 40 to 50 K. No spin
glass phase with vanishing magnetization was found.

The existence of the canted phase requires a careful analysis of a possible
dependence on the cooling process initial conditions. In order to
clarify that point, we performed two additional simulations on sample \#5,
starting at $T=25$ K, where the sample shows already a significant partial
alignment of spins, and we proceeded with the slow cooling down process. In
the first simulation, we assumed a starting configuration in which all spins
are aligned perpendicularly to the interface. In the second, the spins
alignment is made parallel to the interfaces. This choice of a rather low
starting temperature for cooling is necessary, otherwise the thermal
excitation would immediately randomize the initial configuration. The
results are shown in the inset of Fig. \ref{fig01}. We observe that the
appearance of the canted phase does not depend on the choice of the starting
configuration, and the three simulations converge, within statistical
fluctuations, to the same value of the magnetization at every temperature
step.

In Fig. \ref{fig03} the magnetization as a function of temperature is shown
for samples \#08 to \#12, with a higher carrier concentration, $p=0.25 x$,
but keeping the same cutoff radius of 8 ML. The EA order parameter for these
samples shown in Fig. \ref{fig04} as a function of temperature, gives
evidence for the existence of ordered phases. The Fermi wave number
increases with carrier concentration, what decreases the in-plane distance
corresponding to the first zero of the RKKY interaction. In consequence, in
some samples, the magnetic moments order in the canted spin phase (samples
\#10 to \#12). In other samples, however, the ferromagnetic interaction
prevails and the total magnetization is reached (samples \#08 and \#09).
Notice that the canted phase appears here already for $L=50 \AA$, while the
phase is still ferromagnetic for that width when $p=0.1 x$. The transition
temperatures were estimated to lie between 30 K and 50 K.

In Fig. \ref{fig05} we explored the effect of the cutoff radius on the spin
ordering with $p=0.25 x$. A cutoff radius $R_{c}=2$ ML was used in samples
\#13 ($L=60 \AA$) and \#14 ($L=80 \AA$), while $R_{c}=4$ ML was used in
samples \#15 ($L=60 \AA$) and \#16 ($L=80 \AA$). The former is too small,
resulting in the fact that no net magnetization is allowed. The respective
EA order parameters (Fig. \ref{fig06}) are typical of a paramagnetic phases
for samples \#13 and \#14. The choice of a larger $R_{c}$ ordered spins in a
ferromagnetic phase, with transition temperatures calculated to be $35$ K
for the sample \#15, and $80$ K for sample \#16. The cutoff radius of 4 ML
is smaller than the first zero of $J_{ij}^{\text{RKKY}} $. In this situation
the canted spin arrangement is not allowed, and the sample can be either
paramagnetic or ferromagnetic.

Finally in Fig. \ref{fig07} the magnetic susceptibility for sample \#07 is
presented, calculated from the equilibrium magnetization fluctuations.
Notice that the peak in the in the susceptibility indicates the same $T_{c}$
as estimated from the magnetizations and EA order parameter curves (Figs. 
\ref{fig01} and \ref{fig02} respectively).

\section{Discussions and Final Comments}

\label{comments}

The results of the Monte Carlo simulations indicate that, besides the choice
of the range of the interaction (the cutoff radius $R_{c}$), two parameters
are determining in the magnetic ordering in these heterostructures: the
magnetic layer width $L$, and the carrier concentration $p$.

It is observed from resistivity measurements, that the holes have a small
mean free path \cite{matsukura,oiwa} in these materials. The criteria for
choosing the cutoff in a Monte Carlo simulation must take into account
natural scales of the interaction. These scales are the transport mean free
path (since the RKKY interaction is based on the very existence of free
carriers), and also the spin coherence length. We tested different $%
R_{c}$'s, in the range of the mean free path estimated from resistivity
data. The influence of this parameter on the magnetic order is simple. If $%
R_{c}$ is smaller than the first zero of $J_{ij}^{\text{RKKY}}$ (a proper
choice for the case in which the transport mean free path or the spin
coherence length are small), there are two possibilities of magnetic order:
ferromagnetic or paramagnetic. For larger $R_{c}$, on the other hand,
corresponding to the cases where both the transport mean free path and the
coherence length are large, a canted magnetization may be observed.

In all the explored samples, no spin glass phase was found. This is
presumably due to the fact that while the spin frustration exists, as
witnessed by the occurrence of canted spin phases, it is not strong enough
to produce a spin glass phase, as in canonical metallic spin glasses. A spin
glass phase in these DMS structures would probably require a much higher
carrier concentration.

In what concerns the influence of the width of the quantum well, we conclude
that, for thin layers, the number of interacting ions is small within the
cutoff radius, and the sample is paramagnetic. When $L$ becomes larger, the
number of interacting ions increases and a collective magnetic ordering may
be observed. The fact that the appearance of a magnetic order occurs only
above a minimum thickness of the magnetic layer has already been observed
experimentally.\cite{haya}

Since the RKKY interaction oscillates with the argument ($k_{\text{F}}R$),
which depends on the carrier concentration, raising $p$ produces a change on 
$k_{\text{F}}$, increasing the number of oscillations of $J_{ij}^{\text{RKKY}%
}$. Therefore, antiferromagnetic interactions can be turned on, resulting in
all kind of couplings. In this situation, ferromagnetic and
antiferromagnetic interactions compete in establishing the magnetic order,
which depends on the other sample characteristics, resulting in a total or
in a partial alignment of the Mn magnetic moments. The occurrence of partial
magetization (about 40\%) has also been observed in samples of
(In,Mn)As/(Ga,Al)Sb. \cite{oiwa}

To conclude, we believe that the RKKY mechanism explains the high transition
temperatures experimentally observed in Ga$_{1-x}$Mn$_{x}$As
heterostructures, at least in the metallic phase. Additionally, it explains,
as it becomes clear after these Monte Carlo simulations, the occurrence of
samples showing a partial magnetization at low temperatures. The possibility
of having a ferromagnetic phase in samples with a low Mn concentration,
i.e., in the ferromagnetic insulator Ga$_{1-x}$Mn$_{x}$As, remains to be
explained.

\acknowledgements This work was supported by
CENAPAD-SP (Centro Nacional de Processamento de Alto Desempenho em
S\~ao Paulo) UNICAMP/FINEP-MCT, CAPES, CNPq and
FAPERJ in Brazil, and by the PAST grant from Minist\`ere de l'\'Education
Nationale, de l'Enseignement Sup\'erieure et de la Recherche (France).

\newpage 
\begin{table}[tbp]
\caption{Sample characteristics: L is the well width; $r$ is the ratio of
the carrier concentration to the Mn concentration; $T_c$ is the transition
temperature for the phases: F: ferromagnetic, P: paramagnetic, C: canted
spin; $R_c$ is the cutoff radius of the RKKY interaction, in number of
monolayers.}
\label{tab01}
\begin{tabular}{crcccc}
sample & L (\AA) & $r$ & $R_c$ & phase & $T_c$ (K) \\ \hline
\#01 & 50 & 0.01 & 8 & F & 27 \\ 
\#02 & 25 & 0.10 & 8 & P & $\leq$ 1 \\ 
\#03 & 35 & 0.10 & 8 & F & 50 \\ 
\#04 & 50 & 0.10 & 8 & F & 50 \\ 
\#05 & 60 & 0.10 & 8 & C & 50 \\ 
\#06 & 80 & 0.10 & 8 & F & 50 \\ 
\#07 & 100 & 0.10 & 8 & C & 40 \\ 
\#08 & 25 & 0.25 & 8 & F & 50 \\ 
\#09 & 35 & 0.25 & 8 & F & 50 \\ 
\#10 & 50 & 0.25 & 8 & C & 50 \\ 
\#11 & 60 & 0.25 & 8 & C & 40 \\ 
\#12 & 100 & 0.25 & 8 & C & 30 \\ 
\#13 & 60 & 0.25 & 2 & P & $\leq 1$ \\ 
\#14 & 80 & 0.25 & 2 & P & $\leq 1$ \\ 
\#15 & 60 & 0.25 & 4 & F & 35 \\ 
\#16 & 80 & 0.25 & 4 & F & 80
\end{tabular}
\end{table}

\begin{figure}[tbp]
\caption{ Normalized magnetization {\it vs} temperature for samples \#02 to
\#07 indicated in Table {\protect\ref{tab01}}. The inset shows the low temperature
magnetization with three starting spin configurations: random, parallel and
normal to the layers.}
\label{fig01}
\end{figure}

\begin{figure}[tbp]
\caption{ EA parameter {\it vs} temperature for samples \#02 to \#07
indicated in Table {\protect\ref{tab01}}.}
\label{fig02}
\end{figure}

\begin{figure}[tbp]
\caption{ Normalized magnetization {\it vs} temperature for samples \#08 to
\#12 indicated in Table {\protect\ref{tab01}}.}
\label{fig03}
\end{figure}

\begin{figure}[tbp]
\caption{ EA parameter {\it vs} temperature for samples \#08 to \#12
indicated in Table {\protect\ref{tab01}}.}
\label{fig04}
\end{figure}

\begin{figure}[tbp]
\caption{ Normalized magnetization {\it vs} temperature for samples \#13 to
\#16 indicated in Table {\protect\ref{tab01}}.}
\label{fig05}
\end{figure}

\begin{figure}[tbp]
\caption{ EA parameter {\it vs} temperature for samples \#13 to \#16
indicated in Table {\protect\ref{tab01}}.}
\label{fig06}
\end{figure}

\begin{figure}[tbp]
\caption{ Magnetic susceptibility {\it vs} temperature, calculated from
equilibrium magnetization fluctuations, for sample \#07 indicated in Table {%
\protect\ref{tab01}}.}
\label{fig07}
\end{figure}

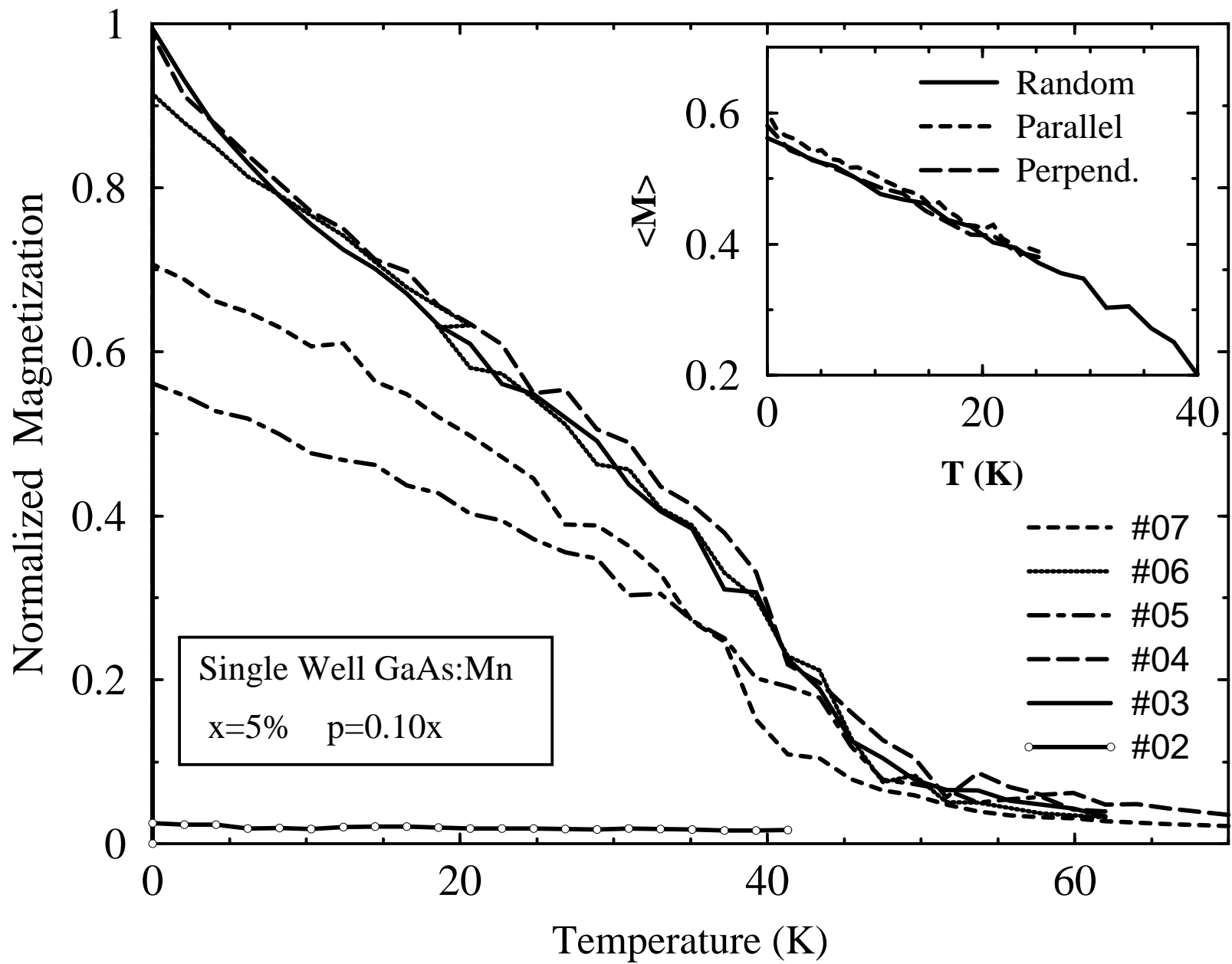
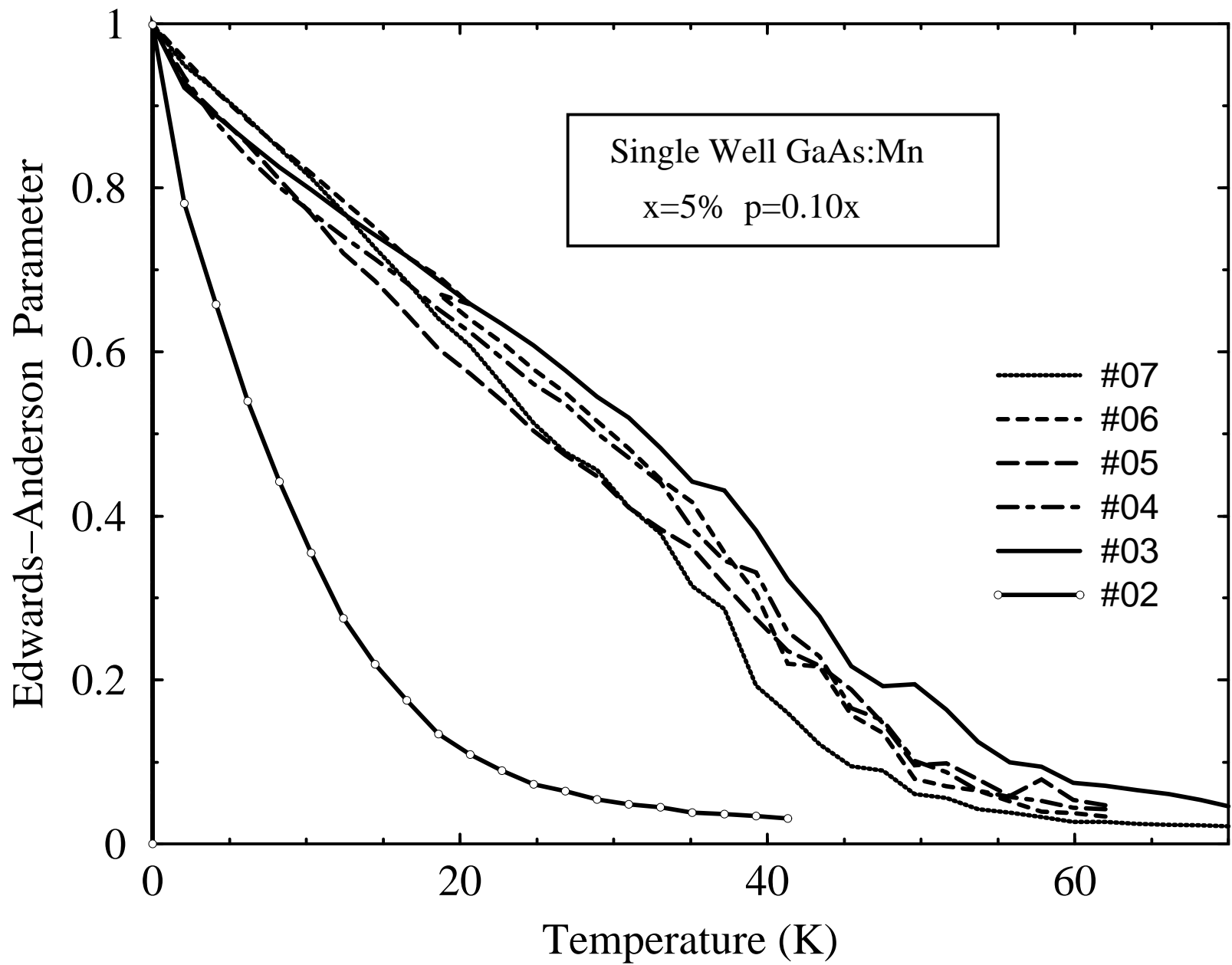
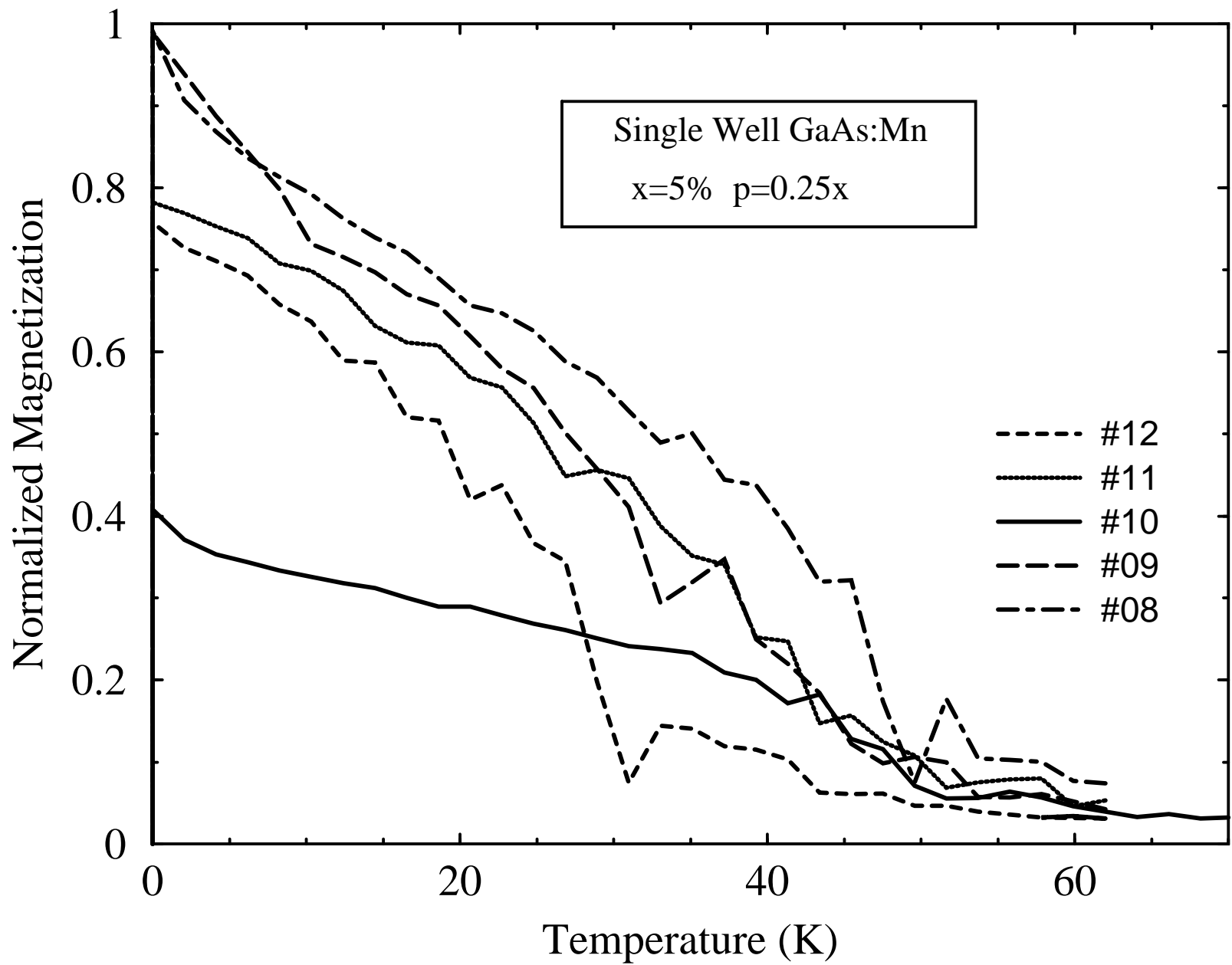
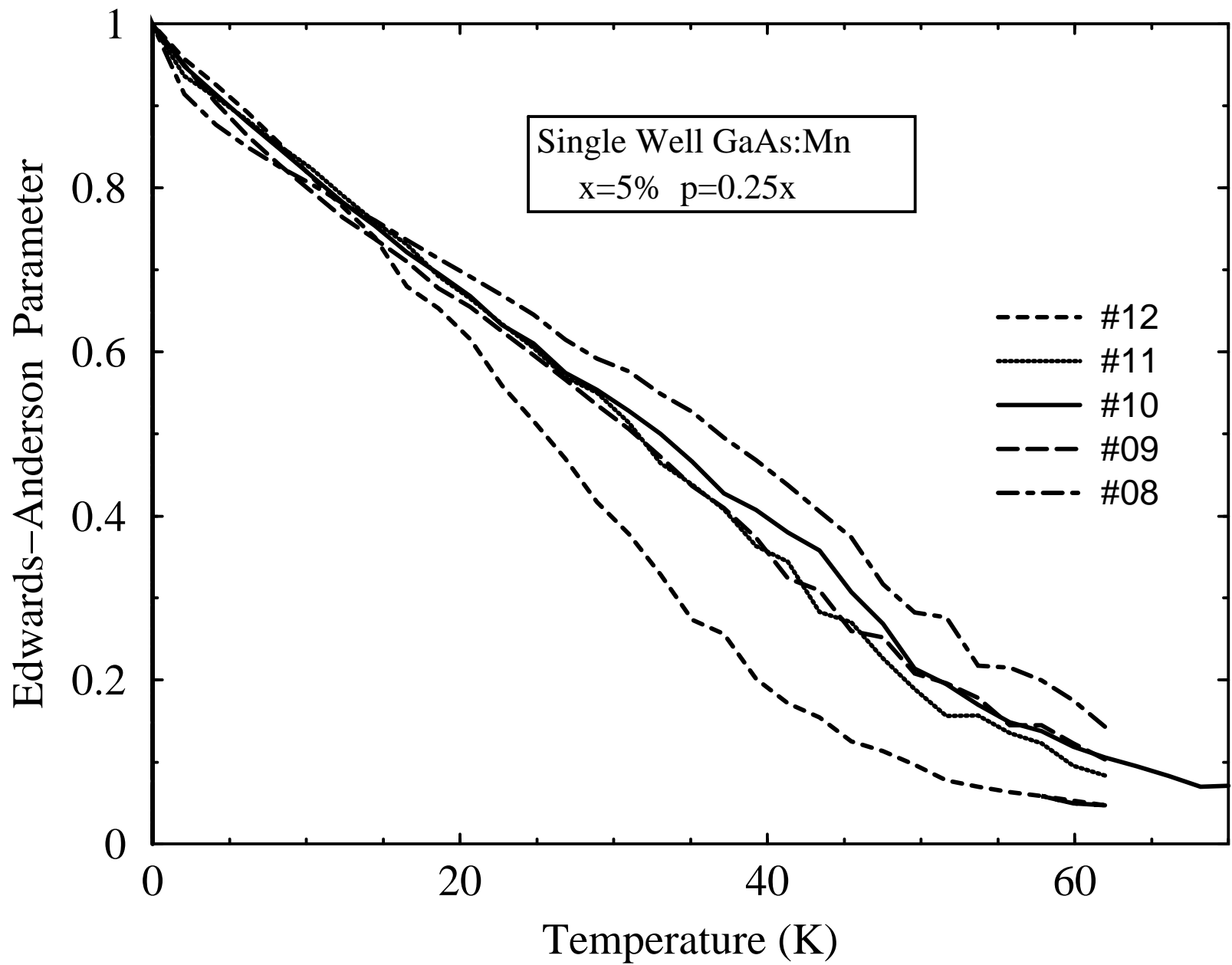
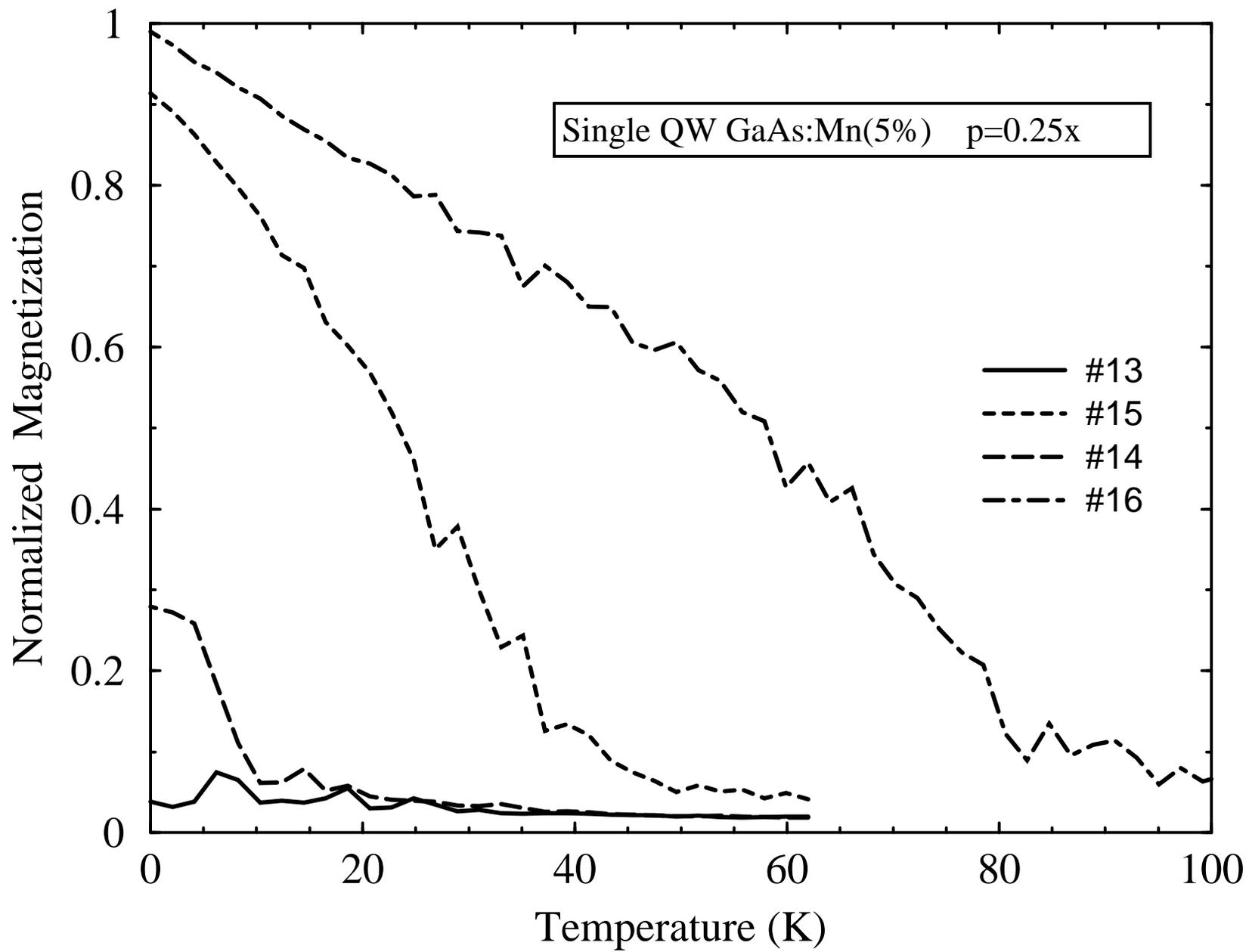
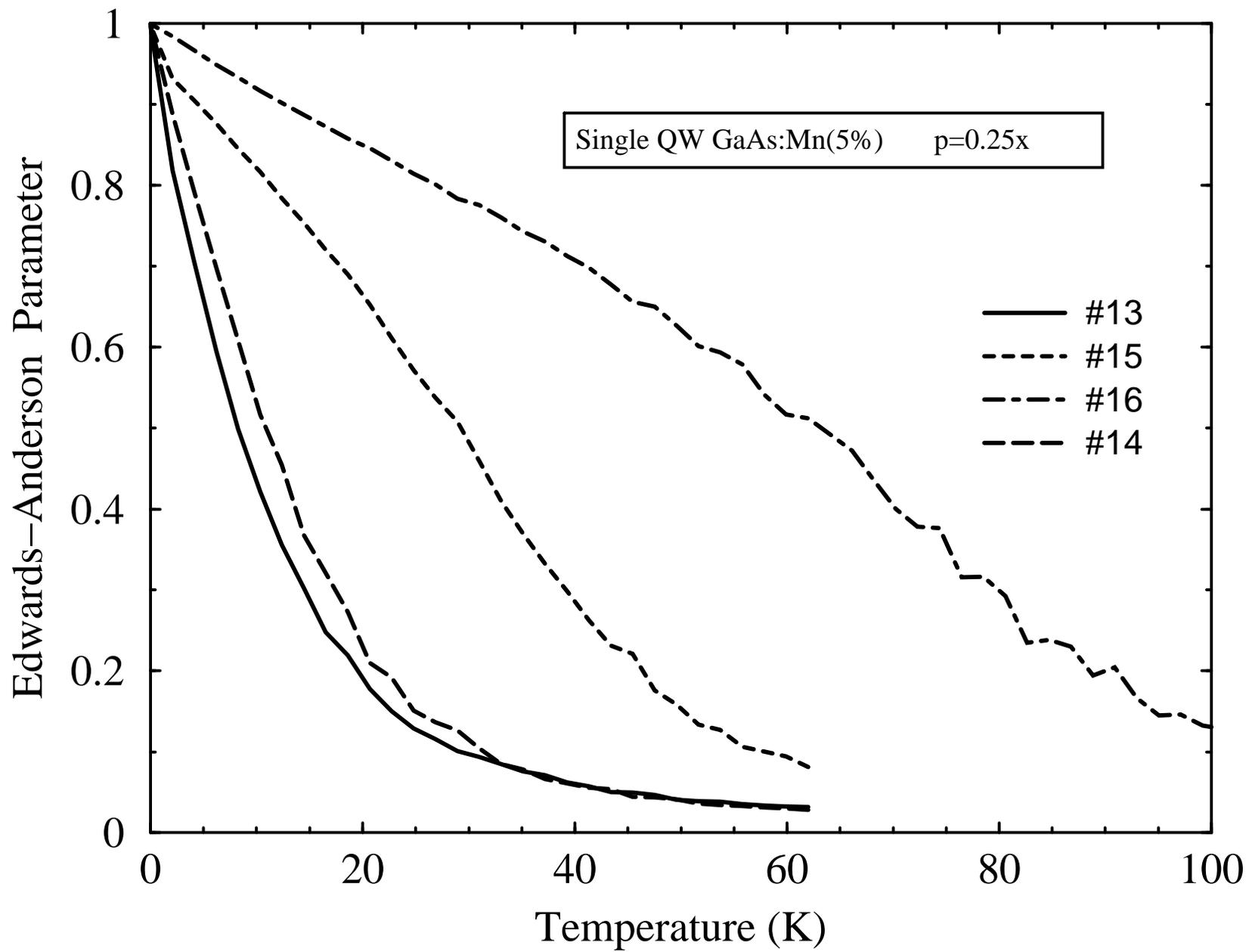
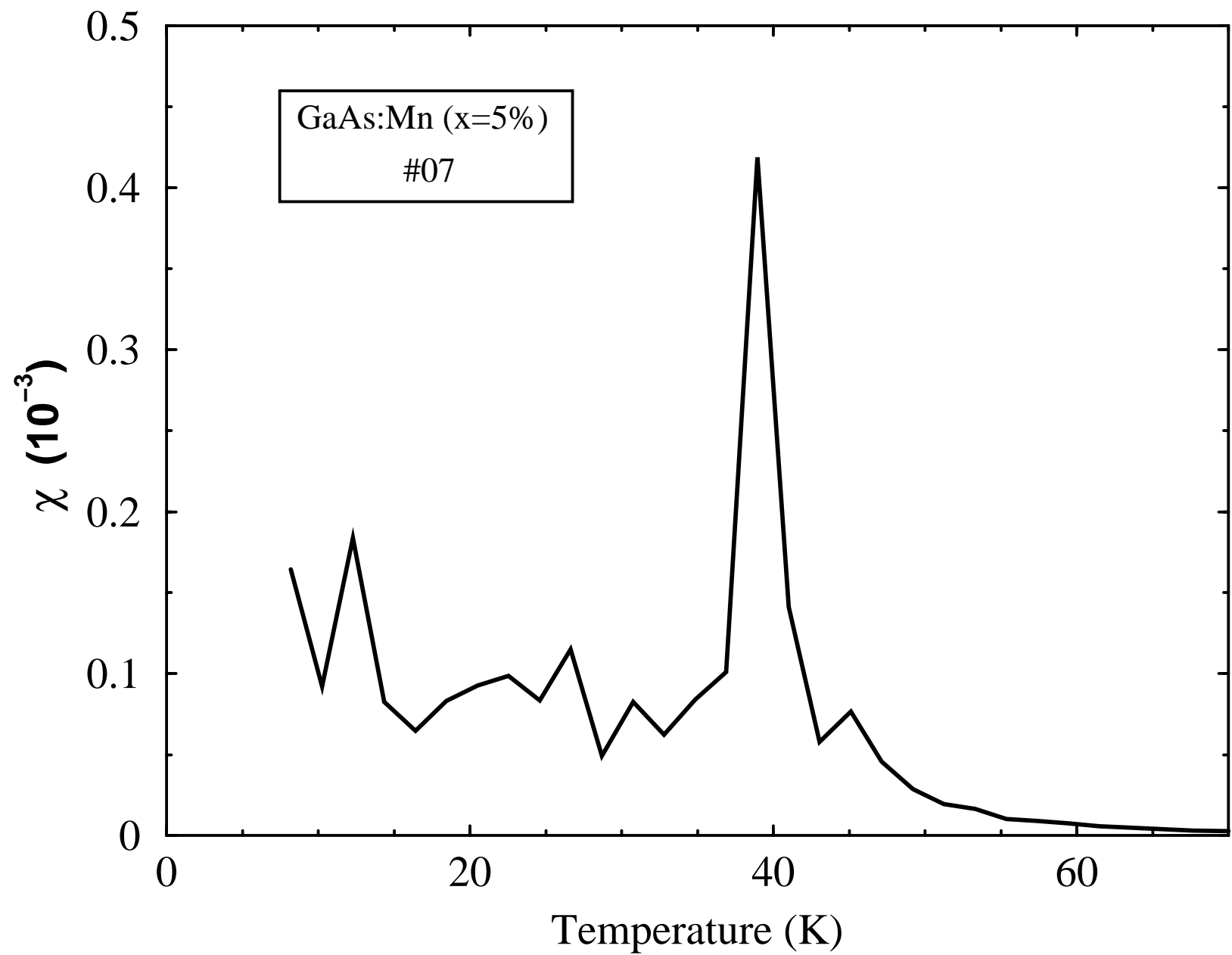
\end{document}